\documentclass[%
preprint,
amsmath,amssymb,
aps,
pre,
]{revtex4-1}

\usepackage{graphicx}
\usepackage{pgf}
\usepackage{pgfplots}
\usepackage{times}
\usepackage{bm}
\usepackage{hyperref}

\newcommand{\ba}[1]{\bm{a}^{(#1)}}
\newcommand{\bb}[1]{\bm{b}^{(#1)}}
\newcommand{\bc}{\bm{c}}
\newcommand{\bdelta}{\bm{\delta}}
\newcommand{\bsigma}{\bm{\sigma}}
\newcommand{\bg}{\bm{g}}
\newcommand{\bk}{\bm{k}}
\newcommand{\bu}{\bm{u}}
\newcommand{\bx}{\bm{x}}
\newcommand{\bxi}{\bm{\xi}}
\newcommand{\bH}[1]{{\cal H}^{(#1)}}
\newcommand{\f}[1]{f^{(#1)}}
\newcommand{\pp}[2]{\frac{\partial #1}{\partial #2}}

\begin{document}

%\begin{CJK*}{UTF8}{} % Use default fonts from CJK (see below)

\title{A central-moment multiple-relaxation-time collision model}
\author{Xiaowen Shan}
\email{shanxw@sustc.edu.cn}
\affiliation{Department of Mechanics and Aerospace Engineering,
  Southern University of Science and Technology, Shenzhen, Guangdong
  518055, China}

\begin{abstract}
  We propose a multiple relaxation time Boltzmann equation collision
  model by systematically assigning a separate relaxation time to each
  of the central moments of the distribution function. The
  Chapman-Enskog calculation leads to correct hydrodynamic
  equations. The thermal diffusion and viscous dissipation are
  mutually independent and Galilean invariant.  By transforming the
  central moments into the absolute reference frame and evaluating
  using fixed discrete velocities, an efficient lattice Boltzmann (LB)
  model is obtained.  The LB model is found to have excellent
  numerical stability in high-Reynolds numbers simulations.
\end{abstract}

\maketitle

%\end{CJK*}

\section{Introduction}

In the past three decades the lattice Boltzmann method
(LBM)~\cite{Frisch1986,Benzi1992,Chen1998a}, particularly the lattice BGK (LBGK)
single-relaxation-time (SRT) model~\cite{Bhatnagar1954,Chen1992,Qian1992}, has
gained a tremendous popularity in many areas of fluid mechanics.  Despite the
great success, a number of deficiencies have long plagued LBGK. The most
noticeable ones are perhaps the fixed unity Prandtl number, the sometimes poor
numerical stability, and the various forms of violations of the Galilean
invariance. Aiming at eliminating, or at least alleviating, these deficiencies,
a number of efforts have been made to improve the collision model, including the
multiple-relaxation-time (MRT) model~\cite{d'Humieres1992,d'Humieres2002} and
its central-moment (CM) version~\cite{Geier2006}, the ``regularized''
models~\cite{Ladd1994a,Latt2006,Chen2006,Mattila2017,Coreixas2017}, and the
Hermite expansion based high-order MRT model~\cite{Shan2007,Chen2014b}.  These
models, suggested with their own purposes and assumptions, all enjoyed success
of various degrees and shared the commonality that the moments of the
distribution are individually manipulated.  One of the aims of the present work
is to offer a coherent view that can hopefully provide a theoretical framework
within which the essence of the aforementioned models can be examined.

The unity Prandtl number is a well-known artifact of the BGK model which relaxes
all moments at the same rate. A few remedies in continuum, \textit{e.g.}, the
ellipsoid-statistical BGK~\cite{Gross1959,Holway1966,Andries2001} and the
Shakhov model~\cite{Shakhov1972} were suggested to introduce additional
parameters in the target distribution so that the heat flux is decoupled from
stress tensor. In the LB realm, McNamara \textit{et al}~\cite{McNamara1997}
implemented a LB collision operator with a different eigenvalue for the third
moments with respect to the \textit{peculiar velocity} to adjust the thermal
conductivity.  In the MRT model of d'Humieres \textit{et
	al}~\cite{d'Humieres2002,Lallemand2003}, the distribution function is decomposed
into eigen-vectors corresponding to the lowest raw moments to each of which a
separate relaxation time is assigned. Theoretically this should allow a variable
Prandtl number. However, as the recovery of the heat equation requires accurate
discrete representation of the third moments~\cite{Nie2008a} which is not
possible on the types of lattices that the MRT was developed with, the MRT was
mostly advocated as a stability improvement.  However, the idea of assigning a
separate relaxation time to each of the moments was generalized to high-order
LBM to allow a variable Prandtl number~\cite{Shan2007}.

In practical simulations, the MRT model was observed to drastically
improve the numerical stability at high Reynolds
numbers~\cite{Lallemand2000,d'Humieres2002}. It is now generally
agreed that this improvement is due to the filtering of the ``ghost''
modes that are not adequately represented by the discrete
velocities~\cite{Nie2008a}. Similar improvements was indeed achieved
by the ``regularized'' models which trim the under-resolved
moments~\cite{Ladd1994a,McNamara1995,Latt2006,Chen2006}. More
recently, the regularization approach was extended to high-order
LB~\cite{Malaspinas2015,Mattila2017,Coreixas2017}, leading to further
enhanced numerical stabilities.

The problem of Galilean invariance has been known since the days of the lattice
Gas Cellular Automaton (LGA) fluid models~\cite{Frisch1986}. Due to
discretization, the hydrodynamic equations differs from the
Navier-Stokes-Fourier (NSF) equations by some velocity-dependent terms. Most of
these problems have been fixed in the LBGK model~\cite{Chen1992,Qian1992} except
the so-called ``cubic'' error~\cite{Qian1993} which results in
velocity-dependent viscosity and/or thermal diffusivity. This is now understood
as being caused by not retaining sufficient moments when the BGK equation is
discretized in velocity space and can be completely eliminated by using
higher-order equilibrium distributions and lattices~\cite{Nie2008a}. Partial
removal of this error is also possible by explicitly correcting the incomplete
third moments~\cite{Dellar2014}. In the high-order MRT model, another violation
of Galilean invariance emerged in the energy equation when the second and third
moments are relaxed at different rates. More recently this error was removed by
explicitly requiring the third moments to take a particular form that yields the
NSF equations~\cite{Chen2014b}, in a similar fashion that the equilibrium
distribution was modified for a similar purpose~\cite{Chen1992}. Nevertheless,
it is not clear how this approach can be extended to the relaxation of
higher-order moments.

The cascaded LB~\cite{Geier2006} (CLB) extends the MRT by performing moment
relaxation in the reference frame moving with the fluid, leading to a cascade of
equations where the relaxations of the higher moments involves those of the
lower ones. Significant improvement of numerical stability has been observed in
simulations~\cite{Ning2016} which is understandable as moment expansion in the
relative frame, \textit{i.e.}, \textit{central moments} (CMs) expansion,
intrinsically has a faster convergence and hence a better numerical performance.
As the derivation of the CLB is rather involved, it is difficult to be extended
beyond the second order to address the unity Prandtl number problem. Also, the
complete restoration of Galilean invariance in the viscous term also requires
correct handling of the third moments~\cite{Nie2008a}.  The viscosity observed
in the simulation~\cite{Geier2006} does show a velocity-dependency despite its
small achievable value.  As far as this author is aware of, there hasn't been
any numerical evidence that CLB has corrected the cubic error.

Lastly, we note that in continuum kinetic theory, moment expansion is
almost always in CMs~\cite{Grad1949a}. However, to compute the CMs
with discrete velocities \textit{via} quadrature, the abscissas must
be chosen in the moving frame and become variables themselves. Sun
\textit{et al}~\cite{Sun2000,Sun2003a} devised an \textit{adaptive} LB
in which the CMs are computed such way and fast convergence were
indeed achieved. The downside is that a complicated particle streaming
scheme involving interpolation has to be adopted. If the advantages of
simple streaming-collision algorithm and linear advection are
preferred, the discrete velocities must be fixed in the absolute
frame.

Here, extending the ideas in Ref.~\cite{Shan2007}, we derive a generic
high-order MRT collision model by separate relaxation of the central moments. We
first note that common to the MRT, regularizations, and cascaded collision
models is the extraction of the moments from the discrete distribution function.
Moments computed by discrete summation are not guaranteed to be the same as the
corresponding velocity integrals in continuum. Namely, the equality:
\begin{equation}
  \label{eq:moments}
  \sum_{i = 1}^d f(\bxi_i)\bxi_i\cdots\bxi_i = \int f(\bxi)\bxi\cdots\bxi d\bxi,
\end{equation}
may or may not hold depending on both the nature of $f$ and the discrete
velocities. In case it doesn't, the hydrodynamic equations of the discrete model
must be re-derived, \textit{e.g.}, by Chapman-Enskog (CE) calculation. Our
formulation here is based on the previous works~\cite{Shan1998,Shan2006b} where
the LB equation was formulated as a special velocity-space discretization of the
continuous BGK equation.  In the classic CE calculation of the BGK
equation~\cite{Huang1987}, the hydrodynamic equations depend only on the leading
CM's of the distribution function instead of its entirety. Provided that the
discrete velocities form a sufficiently accurate quadrature and the equilibrium
distribution is a finite-order truncation of the Maxwellian,
Eq.~(\ref{eq:moments}) is guaranteed up to certain order, and the hydrodynamic
equations are guaranteed to be the same as those obtained from the continuous
BGK equation. The derivation is simple, generic and lattice-independent. The
obtained model has a tunable Prandtl number and Galilean invariant viscous and
thermal dissipations.  At the lowest order, the result of Ref.~\cite{Chen2014b}
is recovered.  In addition, numerical stability similar to or better than those
of the regularized models are achieved.

The work is organized as the following. Theoretical formulation is presented in
Sec.~\ref{sec:LBE}, where, after laying out necessary background, we first
define a transform between the moments and the discrete distribution in
Sec.~\ref{sec:DHT}. In Sec.~\ref{sec:REG} Some previous LB collision models are
examined within this framework.  In Sec.~\ref{sec:MRT} the general conditions
for the collision term to yield NSF equations are obtained by examining the CE
procedure with BGK collision operator~\cite{Chapman1970,Huang1987}. Using these
conditions, a generic high-order MRT collision model is then constructed in
terms of its Hermite expansion. Numerical verifications are provided in
Sec.~\ref{sec:NUM}, and further discussions and conclusions are in
Sec.~\ref{sec:discussion}. Some relations between the moments and Hermite
coefficients in the absolute and relative frames are given in
Appendix~\ref{sec:apdx}.

\section{Theoretical derivation}

\label{sec:LBE}

The LB equation can be viewed as the projection of the following continuous
Boltzmann-BGK equation into a low-dimensional Hilbert space spanned by the
leading Hermite polynomials~\cite{Shan1998,Shan2006b}:
\begin{equation}
  \label{eq:boltzmann}
  \pp ft + \bxi\cdot\nabla f + \bg\cdot\nabla_\xi f = \Omega(f).
\end{equation}
Here, $f$, $\bxi$ and $\bg$ are the single-particle distribution, the peculiar
velocity and the external body force respectively, $\nabla_\xi$ the gradient
operator in velocity space, and $\Omega(f)$ the BGK single-relaxation-time (SRT)
collision model~\cite{Bhatnagar1954}:
\begin{equation}
  \label{eq:bgk}
  \Omega = -\frac 1\tau\left[f - \f{0}\right],
\end{equation}
and $\f{0}$ the Maxwellian:
\begin{equation}
  \label{eq:maxwellian}
  \f{0} = \frac\rho{(2\pi\theta)^{D/2}}
  \exp\left[-\frac{|\bxi - \bu|^2}{2\theta}\right],
\end{equation}
where $\rho$, $\bu$ and $\theta$ are respectively the dimensionless fluid
density, velocity, and temperature~\cite{Shan2006b}.

Hermite polynomials in high dimensions were extensively treated by
Grad~\cite{Grad1949}. Throughout the paper, we use a slightly different notation
which is standard in Tensor Analysis. First define the \textit{symmetrization}
operator:
\begin{equation}
  \mbox{Sym}(\bm{A})\equiv\frac{1}{r!}\sum A_{i_1\cdots i_r},
\end{equation}
where, $\bm{A}$ is a rank-$r$ tensor and the summation is over the $r!$
permutations of the $r$ indexes. The \textit{symmetric product} of two tensors,
$\bm{A}$ and $\bm{B}$, is denoted by $\bm{AB}$ and defined as:
\begin{equation}
  \bm{AB}\equiv\mbox{Sym}(\bm{A}\otimes\bm{B}),
\end{equation}
where $\otimes$ stands for the normal tensor product. The symmetric product has
the following properties:
\begin{enumerate}
\item commutativity: $\bm{AB} = \bm{BA}$;
\item associativity: $(\bm{AB})\bm{C} = \bm{A}(\bm{BC})$;
\item distributivity: $(\bm{A}+\bm{B})\bm{C} = \bm{AC} + \bm{BC}$.
\end{enumerate}
Hereinafter all tensor products are symmetric unless otherwise noted.

\subsection{The discrete Hermite transform}

\label{sec:DHT}

Critical to our formulation of the MRT collision operator is the extraction of
the velocity moments from the \textit{discrete} distributions. For the
hydrodynamic equations to be the NSF equations, we must ensure that the moments
so obtained are exactly the continuum hydrodynamic moments, \textit{i.e.},
Eq.~(\ref{eq:moments}) must hold. Similar to the Discrete Fourier Transforms
(DFT), for a function that is a finite Hermite series, a transform between its
moments and discrete function values can be defined \textit{via} Gauss
quadrature~\cite{Krylov1962,Stroud1971}. First, the Hermite polynomials form an
ortho-normal basis of the $D$-dimensional function space w.r.t.\ the
inter-product $\langle f, g\rangle \equiv\int\omega fgd\bxi$, where
$\omega(\bxi)$ is the weight function:
\begin{equation}
  \frac 1{(2\pi)^{D/2}} \exp\left[-\frac{|\bxi|^2}2\right].
\end{equation}
For any function $f$ such that $f/\omega$ is square-integrable, the following
general Fourier series exists:
\begin{equation}
  \label{eq:dht-a}
  f(\bxi) = \omega(\bxi)\sum_{n=0}^\infty\frac 1{n!}\ba{n}:\bH{n}(\bxi),
\end{equation}
where,
\begin{equation}
  \label{eq:dht-b}
  \ba{n} = \int f(\bxi)\bH{n}(\bxi)d\bxi,\quad n = 0, \cdots, \infty,
\end{equation}
is the $n$-th Hermite coefficients, and `:' denotes full tensor contraction.
Since $\bH{n}(\bxi)$ is a polynomial in $\bxi$, $\ba{n}$ is essentially a
combination of the velocity moments.  For an $N$-th degree polynomials,
$p(\bxi)$, there exists a set of abscissas and associated weights, $\{\bxi_i,
w_i: i = 1, \cdots, d\}$, such that:
\begin{equation}
  \label{eq:quad}
  \int\omega(\bxi) p(\bxi)d\bxi = \sum_{i=1}^dw_ip(\bxi_i).
\end{equation}
Particularly, quadrature rules with abscissas coincide with a Bravais lattice,
\textit{aka} ``on-lattice'' quadratures, can be obtained by solving a
\textit{linear programming}
problem~\cite{Shan2006b,Philippi2006,Chikatamarla2009,Shan2010c,Shan2016}.
Consider the $N$-th order truncation of Eq.~(\ref{eq:dht-a}):
\begin{equation}
  f_N(\bxi) \equiv \omega(\bxi)\sum_{n=0}^N\frac 1{n!}\ba{n}:\bH{n}(\bxi).
\end{equation}
Obviously $f_N/\omega$ is an $N$-th order polynomial.  Eq.~(\ref{eq:dht-b}) can
be written as:
\begin{equation}
  \ba{n} = \int\omega(\bxi)
  \left[\frac{f_N(\bxi)\bH{n}(\bxi)}{\omega(\bxi)}\right]d\bxi.
\end{equation}
Noting that the term inside the brackets is an $(N+n)$-th degree polynomial, by
Eq.~(\ref{eq:quad}) we have:
\begin{equation}
  \label{eq:an}
  \ba{n} = \sum_{i=1}^d f_i\bH{n}(\bxi_i),\quad n = 0, \cdots, Q-N,
\end{equation} 
where $\bxi_i$ and $w_i$ are respectively the abscissas and weights of a
degree-$Q$ quadrature rule, and:
\begin{equation}
  \label{eq:fi}
  f_i \equiv \frac{w_if_N(\bxi_i)}{\omega(\bxi_i)}
  = w_i\sum_{n=0}^N\frac 1{n!}\ba{n}:\bH{n}(\bxi_i).
\end{equation}
Eqs.~(\ref{eq:an}) and (\ref{eq:fi}) define an isomorphic transform between
$\ba{n}$ and $f_i$, allowing the hydrodynamic moments to be exactly computed
from the discrete distribution and \textit{vice versa}.

As shown previously~\cite{Shan2006b}, $f_i$ are exactly the discrete
distribution of LB.  By Eq.~(\ref{eq:an}), the leading moments are the familiar
expressions defining density, $\rho$, velocity, $\bu$, and kinetic energy
density, $\epsilon$:
\begin{equation}
  \rho = \sum_{i=1}^df_i,\quad
  \rho\bu = \sum_{i=1}^df_i\bxi_i,\quad
  \rho(u^2 + 2\epsilon) = \sum_{i=1}^df_i\xi_i^2.
\end{equation}
The dynamic equations for $f_i$ are obtained by directly evaluating
Eq.~(\ref{eq:bgk}) at $\bxi_i$.  After space and time discretization, we can
write the LBGK equation in the following form:
\begin{equation}
  \label{eq:lbgk}
  f_i(\bx + \bxi_i, t+1) = \left(1 - \omega\right)f_i + \omega\f{0}_i,
\end{equation}
where $\omega\equiv 1/\tau$ is the collision frequency. Writing $f_i = \f{0}_i +
\f{1}_i$ with $\f{1}_i$ being the \textit{non-equilibrium} part of the
distribution, the LBGK equation also has the equivalent form:
\begin{equation}
  f_i(\bx + \bxi_i, t+1) = \f{0}_i + \left(1 - \omega\right)\f{1}_i.
\end{equation}

\subsection{Regularization and the general MRT model}

\label{sec:REG}

Eqs.~(\ref{eq:an}) and (\ref{eq:fi}) also provide a natural decomposition of the
discrete distribution, $f_i$, into components corresponding to the moments.
Substituting Eq.~(\ref{eq:an}) into Eq.~(\ref{eq:fi}), we have:
\begin{equation}
  f_i = w_i\sum_{n=0}^N\frac 1{n!}\sum_{j=1}^d\bH{n}(\bxi_j):\bH{n}(\bxi_i)f_j.
\end{equation}
Defining the $d\times d$ \textit{projection matrices}:
\begin{equation}
  \label{eq:m}
  M^{(n)}_{ij} = \frac{w_i}{n!}\bH{n}(\bxi_i):\bH{n}(\bxi_j),
\end{equation}
$M^{(n)}_{ij}f_j$ is the component of $f_i$ corresponding to the $n$-th moment.
Summing up the leading $N$ components, we have the \textit{regularization}
operator:
\begin{equation}
  \label{eq:reg}
  \widehat{f_i} = \sum_{n=0}^NM_{ij}^{(n)}f_j,
\end{equation}
which takes a discrete distribution and trims its Hermite components higher than
$N$. A regularized BGK model similar to that of Ref.~\cite{Latt2006} can then be
generally written as:
\begin{equation}
  \label{eq:lbgk-reg}
  f_i(\bx + \bxi_i, t+1) = \left(1 - \omega\right)\widehat{f_i} + \omega\f{0}_i.
\end{equation}
The collision term on the r.h.s.\ is characterized by three parameters: the
collision frequency, $\omega$, the projection order, $N$, and the truncation
order of $\f{0}$, $M$, which is not necessarily the same as $N$. In case $f_i$
contains no moments beyond the $N$-th order, $\widehat{f_i} = f_i$. Obviously,
when $M \leq N$, we have:
\begin{equation}
  \widehat{\f{0}_i} = \f{0}_i.
\end{equation}
In that case, Eq.~(\ref{eq:lbgk-reg}) can be written as:
\begin{equation}
  f_i(\bx + \bxi_i, t+1) = \f{0}_i + \left(1 - \omega\right)\widehat{\f{1}_i},
\end{equation}
which is essentially an SRT regularized LB that discards all components of the
distribution that correspond to moments higher that what can be accurately
represented by $f_i$.

By assigning a separate relaxation time to each of the Hermite components of
$\f{1}$, the previous MRT LB model~\cite{Shan2007} can be written as:
\begin{equation}
  f_i(\bx + \bxi_i, t+1) = \f{0}_i +
  \sum_{n=2}^N\left(1 - \omega_n\right)M^{(n)}_{ij}\f{1}_j,
\end{equation}
where the summation starts from two because the zeroth and first moments of
$\f{1}$ vanish due to mass and momentum conservation.

\subsection{The multi-relaxation-time collision model}

\label{sec:MRT}

We now turn to the conditions for the collision operator to yield NSF equations
by examining how the NSF equations are derived with the BGK collision
model~\cite{Huang1987}. The hydrodynamic equations are the conservation laws of
mass, momentum and energy, all velocity moments. Taking these moments of
Eq.~(\ref{eq:boltzmann}), the right-hand-side vanishes, and the left-hand-side
contains the following additional central moments:
\begin{equation}
  \bsigma = \int f\bc\bc d\bc,\quad\mbox{and}\quad
  \bm{q} = \frac 12\int f c^2\bc d\bc,
\end{equation}
which are identified as the \textit{pressure tensor} and \textit{heat flux}. We
need to express $\bm{\sigma}$ and $\bm{q}$ in terms of $\rho$, $\bu$, $\theta$
and their derivatives to close the conservation equations. At the zeroth order,
$f$ is taken to be the local Maxwellian of Eq.~(\ref{eq:maxwellian}) which
yields $\bsigma^{(0)} = \rho\theta\bdelta$ and $\bm{q}^{(0)}=\bm{0}$.  On
substituting into the conservation equations, we have Euler's equations. Next,
on substituting $f = \f{0} + \f{1}$ into Eq.~(\ref{eq:boltzmann}) and ignoring
$\f{1}$ on the left-hand-side, we have:
\begin{equation}
  \label{eq:c-moment} 
  \left(\pp{}t + \bxi\cdot\nabla + \bg\cdot\nabla_\xi\right)\f{0} \cong
  -\omega\f{1}.
\end{equation}
The first approximation, $\f{1}$, can be obtained after expressing the l.h.s.\
in terms of $\rho$, $\bu$, $\theta$ and their spatial derivatives by the chain
rule of differentiation and the Euler's equation. Taking the corresponding
moments, we have:
\begin{subequations}
  \label{eq:sigma}
  \begin{eqnarray}
    \sigma^{(1)}_{ij} &=& -\tau\rho\theta\left[\pp{u_i}{x_j} + \pp{u_j}{x_i}-\frac 2D\delta_{ij}\nabla\cdot\bu\right],\\
    \bm{q}^{(1)} &=& -\frac{D+2}2\tau\rho\theta\nabla\theta.
  \end{eqnarray}
\end{subequations}
On substituting the above into the conservation equations we have the NSF
equations.

Evident from this procedure is that the form of the hydrodynamic equations is
completely determined by $\bsigma^{(1)}$ and $\bm{q}^{(1)}$.  As long as the
collision term satisfies the following condition:
\begin{equation}
  \label{eq:omega}
  \int\Omega\bc^nd\bc = -\omega_n\int\f{1}\bc^nd\bc,
  \quad\mbox{for}\quad n = 2, 3.
\end{equation}
$\bsigma^{(1)}$ and $\bm{q}^{(1)}$ will have the same form as
Eqs.~(\ref{eq:sigma}) with $\omega$ replaced by $\omega_2$ and $\omega_3$
respectively.  The hydrodynamic equations will be the same NSF equations but
separately tunable viscosity and thermal diffusivity. More generally, it is
natural to demand that Eq.~(\ref{eq:omega}) is satisfied for all $n$.  This way,
each of the CM's is relaxed at its own rate. Since the set of monomials,
$\{\bc^{n}\}$, is a complete basis of the functional space, by specifying all
moments of $\Omega$, we specify $\Omega$ itself completely.

We now construct the collision operator in terms of its Hermite coefficients.
Let the $n$-th Hermite coefficients of $\Omega$ and $\f{1}$ in absolute frame be
denoted by $\ba{n}_\Omega$ and $\ba{n}_1$ respectively, and those in the
relative frame by $\bb{n}_\Omega$ and $\bb{n}_1$. Due to the conservations of
mass and momentum, we must have $\ba{0}_1 = \ba{1}_1 = 0$, and hence
$\ba{0}_\Omega = \ba{1}_\Omega = 0$. By Eqs.~(\ref{eq:trans-a}), we have
$\bb{0}_1 = \bb{1}_1 = \bb{0}_\Omega = \bb{1}_\Omega = 0$, and:
\begin{subequations}
  \label{eq:trans-s}
  \begin{eqnarray}
    \bb{2} &=& \ba{2},\\
    \bb{3} &=& \ba{3} - 3\bu\ba{2},\\
    \bb{4} &=& \ba{4} - 4\bu\ba{3} + 6\bu^2\ba{2}.
  \end{eqnarray}
\end{subequations}
The Hermite expansions of $\Omega$ and $\f{1}$ in the relative frame are:
\begin{subequations}
  \begin{eqnarray}
    \Omega &=& \omega(\bc)\sum_{n=2}^N\frac 1{n!}\bb{n}_\Omega:\bH{n}(\bc),\\
    \f{1} &=& \omega(\bc)\sum_{n=2}^N\frac 1{n!}\bb{n}_1:\bH{n}(\bc).
  \end{eqnarray}
\end{subequations}
Writing $\bc^n$ in terms of $\bH{n}(\bc)$ by Eqs.~(\ref{eq:hermite-r}) and
using the orthogonality relations, we have:
\begin{subequations}
  \begin{eqnarray}
    \int\Omega\bc^2d\bc &=& \frac{1}{2!}\bb{2}_\Omega,\\
    \int\Omega\bc^3d\bc &=& \frac{1}{3!}\bb{3}_\Omega,\\
    \int\Omega\bc^4d\bc &=& \frac{1}{4!}\bb{3}_\Omega + \frac{6}{2!}\bdelta\bb{2}_\Omega,
  \end{eqnarray}
\end{subequations}
and similar expressions for $\int\f{1}\bc^nd\bc$. On substituting into
Eq.~(\ref{eq:omega}), we arrive at a hierarchy of equations of which the leading
few are:
\begin{subequations}
  \begin{eqnarray}
    \bb{2}_\Omega &=& -\omega_2\bb{2}_1,\\
    \bb{3}_\Omega &=& -\omega_3\bb{3}_1,\\
    \label{eq:b3}
    \bb{4}_\Omega + 72\bdelta\bb{2}_\Omega &=& 
    -\omega_4\left[\bb{4}_1 + 72\bdelta\bb{2}_1\right].
  \end{eqnarray}
\end{subequations}
Converting $\bb{n}$ to $\ba{n}$ using Eq.~(\ref{eq:trans-s}), we have:
\begin{subequations}
  \begin{eqnarray}
    \ba{2}_\Omega &=& -\omega_2\ba{2}_1,\\
    \ba{3}_\Omega &-& 3\bu\ba{2}_\Omega = -\omega_3\left[\ba{3}_1 - 3\bu\ba{2}_1\right],\\
    \ba{4}_\Omega &-& 4\bu\ba{3}_\Omega + 6(\bu^2+12\bdelta)\ba{2}_\Omega = \nonumber\\
    &-&\omega_4\left[\ba{4}_1 - 4\bu\ba{3}_1 + 6(\bu^2+12\bdelta)\ba{2}_1\right].
  \end{eqnarray}
\end{subequations}
Straightforwardly, $\ba{n}_\Omega$ can be solved as:
\begin{subequations}
  \label{eq:mrt}
  \begin{eqnarray}
    \label{eq:mrt-a}
    \ba{2}_\Omega && = -\omega_2\ba{2}_1,\\
    \label{eq:mrt-b}
    \ba{3}_\Omega && = -\omega_3\ba{3}_1 + 3(\omega_3 - \omega_2)\bu\ba{2}_1,\\
    \ba{4}_\Omega && = -\omega_4\ba{4}_1 + 4(\omega_4 - \omega_3)\bu\ba{3}_1 \nonumber\\
    -&& 6[(\omega_4 + \omega_2 - 2\omega_3)\bu^2+12(\omega_4-\omega_2)\bdelta]\ba{2}_1,
  \end{eqnarray}
\end{subequations}
which are the Hermite coefficients of $\Omega$ in the absolute frame. For
comparison, the similar coefficients of the BGK and the high-order
MRT~\cite{Shan2007} operators are respectively:
\begin{equation}
  \label{eq:regnmrt}
  \ba{n}_\Omega = -\omega\ba{n}_1,\quad\mbox{and}\quad
  \ba{n}_\Omega = -\omega_n\ba{n}_1.
\end{equation}

We first note that when all the relaxation times are the same, all three are
identical.  Second, as far as the second moments are concerned, relaxations of
the central and raw moments are equivalent. This is in agreement with some of
the numerical observation~\cite{Ning2016}. Third, the correction to the third
moments, \textit{i.e.}, the second term on the right-hand-side of
Eq.~(\ref{eq:mrt-b}), recovers the result in Ref.~\cite{Chen2014b}.

The computation of the collision process goes as the following. Given the
post-streaming distribution, $f_i$, its non-equilibrium part is $\f{1}_i = f_i -
\f{0}_i$, from which $\ba{n}_1$, and $\ba{n}_\Omega$ in turn, can be calculated
by Eqs.~(\ref{eq:an}) and (\ref{eq:mrt}). $\Omega_i$ is then obtained from
$\ba{n}_\Omega$ using Eq.~(\ref{eq:fi}), and finally the post-collision
distribution is updated using the following lattice Boltzmann equation:
\begin{equation}
  f_i(\bx + \bxi_i, t+1) = \hat{f_i} + \Omega_i.
\end{equation}

\section{Numerical verification}

\label{sec:NUM}

In this section we numerically verify the CM-based MRT (CM-MRT) model. First the
viscosity and thermal diffusivity were numerically measured \textit{via} the
dynamics of the linear hydrodynamic modes in the presence of a translational
flow. The numerical measurements are then compared with theoretical values.  The
independence of the transport coefficients on the translational flow, and hence
the Galilean invariance in the dissipation terms, can then be verified.
Secondly, a thorough and complete characterization of CM-MRT's numerical
stability is beyond the scope of the present paper and deferred to a later
publication. Here we choose to only present some preliminary results on the
popular test case of the double shear layer~\cite{Brown1995,Minion1997}.  The
results seem to show that the CM-MRT is at least as stable as the regularized
collision models.

\subsection{Linear hydrodynamic modes test}

\newcommand{\pe}{\mbox{Pe}}
\newcommand{\re}{\mbox{Re}}
\newcommand{\pr}{\mbox{Pr}}
\newcommand{\ma}{\mbox{Ma}}

We first give the theoretical predictions of the viscous, thermal and acoustic
modes in the presence of a translational flow.  Consider a small perturbations
on top of a base flow with constant velocity. The density, velocity and
temperature, all non-dimensionalized by the scheme in Ref.~\cite{Shan2006b}, are
written as:
\begin{equation}
  \label{eq:32}
  \rho = \rho_0 + \rho',\quad
  \bu  = \bu_0 + \bu',\quad\mbox{and}\quad
  \theta = \theta_0 + \theta'.
\end{equation}
where the subscript $_0$ and the prices denote the quantities of the base flow
and the perturbation respectively. The perturbation is in the form of a
monochromatic wave:
\begin{equation}
  \label{eq:modes}
  \left(\begin{array}{c}\rho'\\ \bu'\\ \theta' \end{array}\right) =
  \left(
  \begin{array}{c}\tilde{\rho}\\ \tilde{\bu}\\ \tilde{\theta} \end{array}
  \right)e^{\omega t+i\bm{k}\cdot(\bx - \bu_0t)},
\end{equation}
where $\tilde{\rho}$, $\tilde{\bu}$ and $\tilde{\theta}$ are constant scaler
amplitudes of the perturbations, $\omega$ and $\bk$ the frequency and wave
vector, and $\bx$ the spatial coordinate. We first decompose the velocity
perturbation into components parallel and perpendicular to the wave vector,
\textit{i.e.}, we write $\tilde{\bu} = \tilde{u}_\parallel\bm{e}_\parallel +
\tilde{u}_\perp\bm{e}_\perp$, where $\bm{e}_\parallel$ and $\bm{e}_\perp$ are
unit vectors parallel and perpendicular to $\bk$.  On substituting
Eqs.~(\ref{eq:32}) and (\ref{eq:modes}) into the NSF equations, we obtain an
eigen-system in the linear space of $(\tilde{\rho}, \tilde{u}_\parallel,
\tilde{\theta}, \tilde{u}_\perp)^T$. The four eigen values give the dispersion
relations, while the eigen-vectors define the corresponding amplitudes.

Let $\gamma$ be the \textit{heat capacity ratio}, $\nu$ and $\eta$ the first and
second \textit{kinematic viscosities}, and $\kappa$ the thermal diffusivity.
Further non-dimensionalizing by defining the \textit{acoustic} Reynolds and
P\'eclet numbers as $\re = c_s/\nu k$ and $\pe = c_s/\kappa k$, where $c_s\equiv
\sqrt{\gamma\theta_0}$ is the \textit{sound speed} at temperature $\theta_0$.
$\re$ and $\pe$ are related by $\pe = \re\cdot\pr$ where $\pr\equiv\nu/\kappa$
is the Prandtl number. The dimensionless dispersion relations are:
\begin{subequations}
  \label{eq:dr}
  \begin{eqnarray}
    -\frac{\omega_v}{c_sk} &=& \frac 1{\re},\\
    -\frac{\omega_t}{c_sk} &=& \frac 1{\pe} + \frac{(\gamma-1)\lambda}{\pe^3}
    + \mathcal{O}\left(\frac{1}{\pe^5}\right),\\
    -\frac{\omega_\pm}{c_sk} &=& \frac{\gamma -\lambda}{2\pe} -\frac{(\gamma-1)\lambda}{2\pe^3} +  \mathcal{O}\left(\frac{1}{\pe^5}\right)\nonumber\\
    &\pm& i\left[1-\frac{(\gamma +\lambda)^2 - 4\lambda}{8\pe^2}
      + \mathcal{O}\left(\frac{1}{\pe^4}\right)\right],
  \end{eqnarray}
\end{subequations}
where $\omega_v$, $\omega_t$, and $\omega_\pm$ are the angular frequencies of
the viscous, thermal, and acoustic modes respectively, $\lambda\equiv 1 +
(\gamma-3)\Pr$ is a constant defined for brevity.  Note that while the
dispersion relation of the viscous mode is exact, the other three are solutions
of a cubic characteristic equation and only their asymptotic expansions at the
small-Pe limit are given.

The numerical measurements were carried out as the following.  First, given the
desired amplitudes of the four modes, $\tilde{\rho}$, $\tilde{\bu}$ and
$\tilde{\theta}$ were determined as the superposition of the four eigen-vectors.
The perturbations, $\rho'$, $\bu'$ and $\theta'$, were then constructed
according to Eqs.~(\ref{eq:modes}).  Subsequently the amplitudes were determined
by performing a spatial fast Fourier transform on a corresponding data field to
extract the component of the given wave number. Noting that sound propagation is
isentropic and thermal diffusion is isobaric, the data field for the viscous,
thermal, and acoustic modes is $u_\perp$, the pressure, $p\equiv\rho\theta$, and
entropy, $s\equiv c_v\ln(\theta\rho^{1-\gamma})$, respectively.

\begin{figure}
  \centering
  \begin{tikzpicture}
    \begin{semilogyaxis}[
	ytickten={-1,0},
	yticklabels={0.1, 1},
	enlarge x limits=false,
	ymin=0.08,
	xlabel=Time,
	ylabel={Normalized amplitudes},
	legend pos=south east,
	grid=major]
      \addplot[brown,mark=triangle,only marks] table[x=t, y=vis] {modes.dat};
      \addplot[red,  mark=square,  only marks] table[x=t, y=thm] {modes.dat};
      \addplot[blue, mark=o,       only marks] table[x=t, y=acs] {modes.dat};
      \addplot[domain=0:350,y domain=0.08:1, samples=500] {exp(-0.0004133*x)*abs(cos(0.0813*deg(x)))};
      \addplot[domain=0:350] {exp(-0.0002755*x)};
      \addplot[domain=0:350] {exp(-0.0011*x)};
      \legend{Viscous mode,Thermal mode, Acoustic mode}
    \end{semilogyaxis}
  \end{tikzpicture}
  \caption{Typical time histories of the linear mode amplitudes.  The
    simulation was performed using the 2D 37-velocity $E^{37}_{2,9}$
    quadrature~\cite{Shan2016} on a $100\times 100$ double periodic
    lattice.  Shown are the absolute values of the amplitudes of the
    viscous, thermal and standing acoustic waves all normalized with
    their initial values.  The solid lines are theoretical results and
    symbols numerical measurements.}
  \label{fg:amp}
\end{figure}
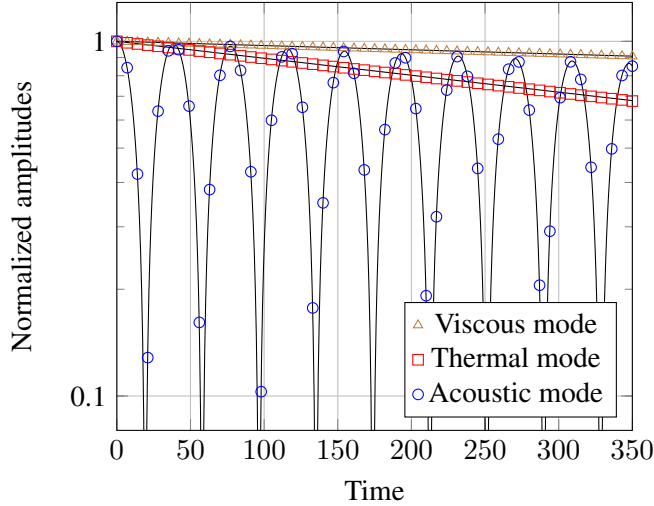

Shown in Fig.~\ref{fg:amp} is the typical behavior of the linear mode amplitudes
against their theoretical values.  The CM-MRT model with a ninth-order 37-speed
quadrature is used. The simulation was performed with $\nu = 0.1$ and $\kappa =
0.2$, yielding a Prandtl number of 0.5. The density, temperature and
translational velocity of the base flow are $\rho_0 = 1$, $\theta_0 = 1.2$ and
$\bu_0 = \bm{0}$ with the initial perturbation being a superposition of three
monochrome viscous, thermal and standing acoustic wave, all with amplitude 0.001
and wave number $(1, 0)$, $(1,1)$ and $(1,0)$ respectively. The time histories
were then fitted with the theoretical model of Eq.~(\ref{eq:modes}) to determine
the angular frequencies. Comparing with Eqs.~(\ref{eq:dr}), the errors in
$\omega_v$, $\omega_t$ and the real and imaginary parts of $\omega_\pm$ are
respectively 0.17\%, 0.19\%, 0.19\% and 0.01\%.

\begin{figure}
  \centering
  \begin{tikzpicture}
    \begin{loglogaxis}[
	ylabel=Relative error in viscosity,
	grid=major,
	xtick={20, 40, 80, 160, 320},
	xticklabels=\empty
      ]
      \addplot[mark=o]        table[x index=0, y index=1] {convergence.dat};
      \addplot[mark=x]        table[x index=0, y index=3] {convergence.dat};
      \addplot[mark=square]   table[x index=0, y index=5] {convergence.dat};
      \addplot[mark=triangle] table[x index=0, y index=7] {convergence.dat};
      \addplot[red,mark=o]        table[x index=0, y index=9] {convergence.dat};
      \addplot[red,mark=x]        table[x index=0, y index=11] {convergence.dat};
      \addplot[red,mark=square]   table[x index=0, y index=13] {convergence.dat};
      \addplot[red,mark=triangle] table[x index=0, y index=15] {convergence.dat};
      \legend{
	{$E^7_{2,17}$-A},
	{$E^7_{2,17}$-B},
	{$E^7_{2,17}$-C},
	{$E^7_{2,17}$-D},
	{$E^9_{2,37}$-A},
	{$E^9_{2,37}$-B},
	{$E^9_{2,37}$-C},
	{$E^9_{2,37}$-D}
      }
    \end{loglogaxis}
  \end{tikzpicture}
  \begin{tikzpicture}
    \begin{loglogaxis}[
	xlabel=Grid resolution,
	ylabel=Relative error in thermal diffusivity,
	grid=major,
	xtick={20, 40, 80, 160, 320},
	xticklabels={20, 40, 80, 160, 320}
      ]
      \addplot[mark=o]        table[x index=0, y index=2] {convergence.dat};
      \addplot[mark=x]        table[x index=0, y index=3] {convergence.dat};
      \addplot[mark=square]   table[x index=0, y index=4] {convergence.dat};
      \addplot[mark=triangle] table[x index=0, y index=5] {convergence.dat};
      \addplot[red,mark=o]        table[x index=0, y index=10] {convergence.dat};
      \addplot[red,mark=x]        table[x index=0, y index=12] {convergence.dat};
      \addplot[red,mark=square]   table[x index=0, y index=13] {convergence.dat};
      \addplot[red,mark=triangle] table[x index=0, y index=14] {convergence.dat};
      \legend{
	{$E^7_{2,17}$-A},
	{$E^7_{2,17}$-B},
	{$E^7_{2,17}$-C},
	{$E^7_{2,17}$-D},
	{$E^9_{2,37}$-A},
	{$E^9_{2,37}$-B},
	{$E^9_{2,37}$-C},
	{$E^9_{2,37}$-D}
      }
    \end{loglogaxis}
  \end{tikzpicture}
  \caption{Grid convergence of the CM-MRT model.  Plotted are the
    relative errors in viscosity (top) and thermal diffusivity
    (bottom) using the $E^7_{2,17}$ and $E^9_{2,37}$ quadratures on a
    $L\times L$ lattice ranging from $L=20$ to $L=320$.  The errors in
    viscosity of the four $E^9_{2,37}$ quadrature are almost identical
    and coincide on the graph.  Although all models are second order,
    the magnitudes of the error can differ by a factor of
    approximately 4-5 among all quadrature rules.  The quadrature
    $E^7_{2,17}$-D is found to have the best
    accuracy.}
  \label{fg:grid}
\end{figure}
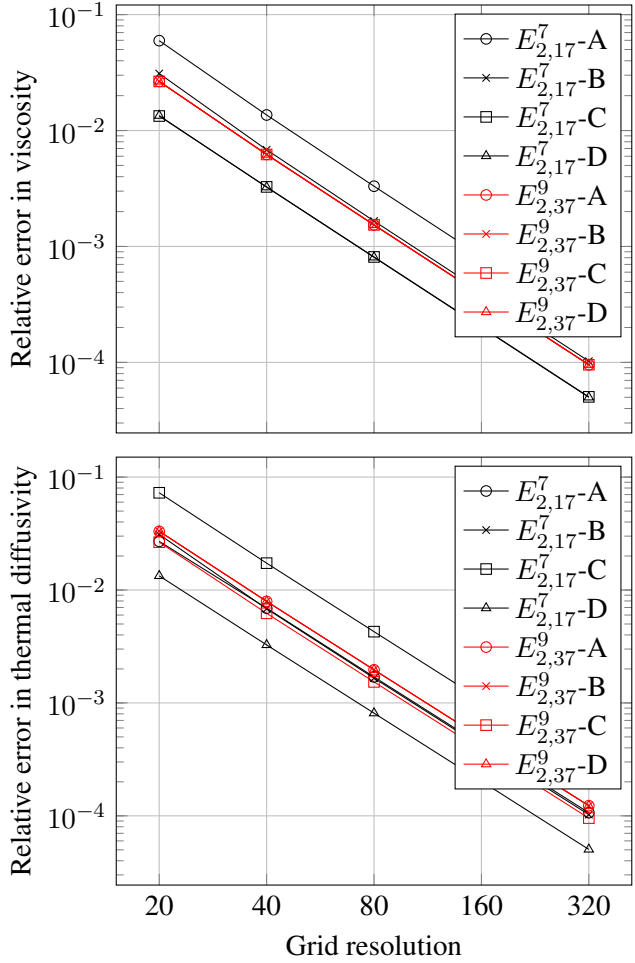

Using this measurement mechanism, we first tested the grid convergence of the
CM-MRT model with a number of high-order quadratures.  As previously
shown~\cite{Shan2016}, high-order quadrature rules with abscissas coincide with
lattice nodes (on-lattice) that can accurately represent moments of any order
can be found by solving a linear programming problem.  The solutions form an
polytope in the parameter space with its vertexes representing the quadratures
with minimum number of velocities.  In 2D, the minimum 7-th degree quadrature
rules are the four $E^7_{2,17}$ rules, and the minimum 9-th degree rules are the
four $E^9_{3,37}$ rules, all given in Ref.~\cite{Shan2016}.  Shown in
Fig.~\ref{fg:grid} are the relative errors in viscosity and thermal diffusivity
measured by linear mode simulations using the CM-MRT model with all eight
quadratures. All models demonstrate a second order spatial accuracy with the one
using quadrature $E^7_{2,17}$-D being the most accurate.

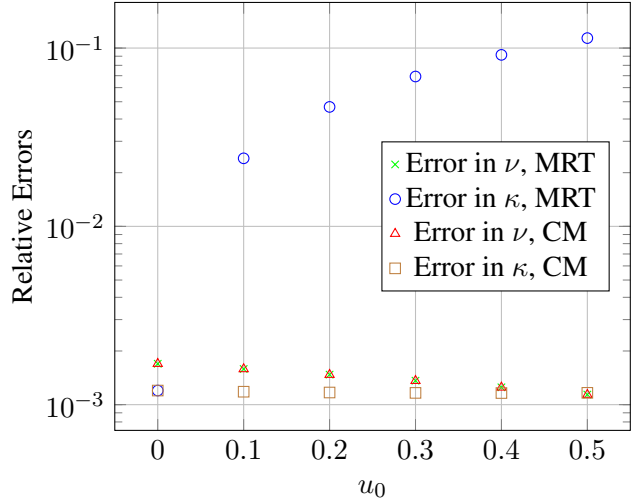
\begin{figure}
  \centering
  \begin{tikzpicture}
    \begin{semilogyaxis}[
	xlabel=$u_0$,
	ylabel={Relative Errors},
	legend style={at={(0.96,0.5)}, anchor=east},
	grid=major]
      \addplot[green,mark=x,only marks] table[x=u0, y=evm] {errors.dat};
      \addplot[blue,mark=o,only marks] table[x=u0, y=etm] {errors.dat};
      \addplot[red,mark=triangle,only marks] table[x=u0, y=evc] {errors.dat};
      \addplot[brown,mark=square,only marks] table[x=u0, y=etc] {errors.dat};
      \legend{
	{Error in $\nu$, MRT},
	{Error in $\kappa$, MRT},
	{Error in $\nu$, CM},
	{Error in $\kappa$, CM}
      }
    \end{semilogyaxis}
  \end{tikzpicture}
  \caption{Restoration of the Galilean invariance of transport
    coefficients by the CM-MRT model.  Plotted are the relative errors
    in viscosity, $\nu$, and thermal diffusivity, $\kappa$, as
    measured from the linear mode tests using MRT and CM-MRT models,
    both with the $E^{37}_{2,9}$ quadrature on a $100\times 100$
    lattice.  On the horizontal axis is the magnitude of the
    translational velocity.  The error in thermal diffusivity in the
    MRT model increases with $u_0$, breaking the Galilean
    invariance.}
  \label{fg:galilean}
\end{figure}

We then used this apparatus to verify the Galilean invariance by including a
translational velocity in the base flow, in a similar fashion as in
Ref.~\cite{Chen2014b}. Specifically we set $\bu_0 = (0, u_0)$, and the initial
perturbation consists of a viscous and a thermal wave, both with wave vector
$\bk = (1, 0)$ and initial amplitude of 0.001.  The base flow is in the
transverse direction of the wave vector.  Shown in Fig.~\ref{fg:galilean} are
the errors in the measured viscosity and thermal diffusivity against $u_0$ using
the MRT~\cite{Shan2007} and CM-MRT models.  To be seen is that the errors in
viscosity are small and identical, confirming the theoretical finding that the
relaxations of raw and central moments at the second order are identical. The
error in thermal diffusivity in MRT however increases linearly with $u_0$. This
violation of Galilean invariance is eliminated in CM-MRT.

\subsection{Double shear layer test}

\begin{figure}
  \begin{tikzpicture}
    \pgfplotsset{
      axis x line=none,
      axis y line=none,
      width=2.2in,
      height=2.2in,
      enlarge x limits=0.01,
      enlarge y limits=0.01
    }
    \matrix {
      \begin{axis}
	\addplot graphics[xmin=0,xmax=1,ymin=0,ymax=1]{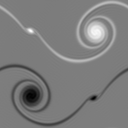};
      \end{axis}
      &
      \begin{axis}
	\addplot graphics[xmin=0,xmax=1,ymin=0,ymax=1]{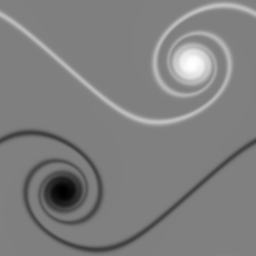};
      \end{axis}
      \\
    };
  \end{tikzpicture}
  \caption{Vorticity field at $t = 1$ in a double shear layer
    simulation using the D2Q9 BGK model.  On the left, the simulation
    resolution is $128\times 128$, deemed insufficient as indicated by
    the spurious secondary vortexes that are absent in the better
    resolved case ($256\times 256$) on the right. The Reynolds number
    and Mach number are respectively 10,000 and 0.1 in both cases.}
  \label{fg:dsl}
\end{figure}

The double-shear-layer (DSL)~\cite{Brown1995,Minion1997} is a well
studied test case for numerical stability benchmark~\cite{Dellar2001,
  Dellar2003, Bosch2015, Mattila2015, Mattila2017, Coreixas2017}. The
two-dimensional flow field is defined on a double periodic domain $0
\leq x, y \leq 1$ by:
\begin{subequations}
\begin{eqnarray}
  u_x &=& \left\{
  \begin{array}{ll}
    u_0\tanh\rho(y-\frac 14), & \quad y \leq\frac 12\\
    u_0\tanh\rho(\frac 34-y), & \quad y > \frac 12
  \end{array},
  \right.\\
  u_y &=& \delta u_0\sin 2\pi\left[x + \frac 14\right],
\end{eqnarray}
\end{subequations}
where $1/\rho$ measures the thickness of the shear layer, and $\delta$ a small
parameter controlling the magnitude of the initial vertical perturbation.  In
simulations here, we chose $\rho = 80$ and $\delta = 0.05$ in accordance with
the literature.  All simulations are performed on a $L\times L$ square lattice
where $L$ is the number of sites in one direction. In our
notation~\cite{Shan2006b}, lengths are scaled by the lattice constant, $c$, and
velocities by the isothermal sound speed, $c_s$. The Reynolds and Mach numbers
are therefore $\re = u_0cL/\nu$ and $\ma = u_0$. Shown in Fig.~\ref{fg:dsl} are
the typical vorticity fields simulated using D2Q9 with the resolutions $L=128$
and $L=256$ respectively. The occurrence of the secondary vortexes on the left
is a well-known indication of insufficient resolution.

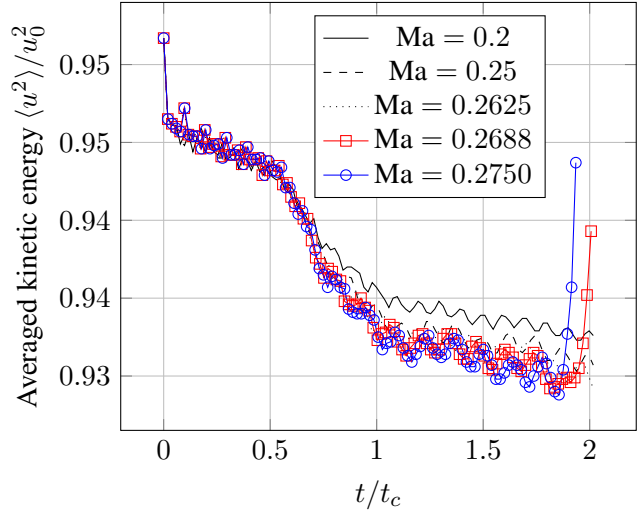
\begin{figure}
  \centering
  \begin{tikzpicture}
    \begin{axis}[
	xlabel=$t/t_c$,
	ylabel={Averaged kinetic energy $\langle u^2\rangle/u_0^2$},
	legend style={at={(0.6,0.97)}, anchor=north},
	grid=major]
      \addplot[mark=none,style=solid] table
      	{D2Q37A_CM4_4_THM_RE_1e7_MA_0_2000_PR_1_000_128.dat};
      \addplot[mark=none,style=dashed] table
       {D2Q37A_CM4_4_THM_RE_1e7_MA_0_2500_PR_1_000_128.dat};
      \addplot[mark=none,style=dotted] table
       {D2Q37A_CM4_4_THM_RE_1e7_MA_0_2625_PR_1_000_128.dat};
      \addplot[mark=square,style=solid,red] table
       {D2Q37A_CM4_4_THM_RE_1e7_MA_0_2688_PR_1_000_128.dat};
      \addplot[mark=o,style=solid,blue] table
       {D2Q37A_CM4_4_THM_RE_1e7_MA_0_2750_PR_1_000_128.dat};
      \legend{
	{$\ma=0.2$},
	{$\ma=0.25$},
	{$\ma=0.2625$},
	{$\ma=0.2688$},
	{$\ma=0.2750$}
      }
    \end{axis}
  \end{tikzpicture}
  \caption{(Color online) Time histories of the averaged kinetic energy
	normalized by its initial value for a sequence of Mach number at $\pr=1$ and
	$\re=10^7$ using thermal CM-MRT on an $128\times 128$ lattice. At $\ma=0.275$
	the simulation diverged and at $0.2688$ it barely survived beyond $t/t_c =
	2$.} 
  \label{fg:converg}
\end{figure}

Extensive studies on the DSL were carried out to benchmark various collision
models~\cite{Mattila2017,Coreixas2017}.  For comparison, we also computed the
stability boundary of the DSL using the isothermal MRT, isothermal CM-MRT and
full thermal CM-MRT models on the same $L=128$ lattice. For a fixed pair of
$\pr$ and $\re$, the maximum $\ma$ is defined as the highest $\ma$ that allows
the simulation to be stably carried out till $t/t_c = 2$~\cite{Coreixas2017}. An
iterative search algorithm was used to found the maximum $\ma$ for fixed $\pr$
and $\re$. Shown in Fig.~\ref{fg:converg} are the time histories of the averaged
kinetic energy, $\langle u^2\rangle/u_0^2$, for an increasing sequence of Mach
numbers at $\pr = 1$ and $\re = 10^7$ using the thermal CM-MRT.  The maximum
$\ma$ is determined at $0.2688$ in this case.

\begin{figure}
  \centering
  \begin{tikzpicture}
    \begin{semilogxaxis}[
	ymin=0.5,ymax=0.8,
	ylabel={Maximum Mach number},
	legend pos=south east,
	xticklabels=\empty,
	grid=major
      ]
      \addplot table[x index=0,y index=1] {D2Q37A_MRT4_4_ISO_stability.dat};
      \addplot table[x index=0,y index=2] {D2Q37A_MRT4_4_ISO_stability.dat};
      \addplot[mark=o] table[x index=0,y index=3] {D2Q37A_MRT4_4_ISO_stability.dat};
      \addplot[mark=square] table[x index=0,y index=4] {D2Q37A_MRT4_4_ISO_stability.dat};
      \addplot[mark=x] table[x index=0,y index=5] {D2Q37A_MRT4_4_ISO_stability.dat};
      \legend{
	{Pr = 0.01},
	{Pr = 0.1},
	{Pr = 1.},
	{Pr = 10.},
	{Pr = 100.}
      }
    \end{semilogxaxis}
  \end{tikzpicture}
  \begin{tikzpicture}
    \begin{semilogxaxis}[
	ymin=0.5,ymax=0.8,
	ylabel={Maximum Mach number},
	legend pos=south east,
	xticklabels=\empty,
	grid=major
      ]
      \addplot table[x index=0,y index=1] {D2Q37A_CM4_4_ISO_stability.dat};
      \addplot table[x index=0,y index=2] {D2Q37A_CM4_4_ISO_stability.dat};
      \addplot[mark=o] table[x index=0,y index=3] {D2Q37A_CM4_4_ISO_stability.dat};
      \addplot[mark=square] table[x index=0,y index=4] {D2Q37A_CM4_4_ISO_stability.dat};
      \addplot[mark=x] table[x index=0,y index=5] {D2Q37A_CM4_4_ISO_stability.dat};
      \legend{
	{Pr = 0.01},
	{Pr = 0.1},
	{Pr = 1.},
	{Pr = 10.},
	{Pr = 100.}
      }
    \end{semilogxaxis}
  \end{tikzpicture}
  \begin{tikzpicture}
    \begin{semilogxaxis}[
	ymin=0.2,ymax=0.8,
	xlabel=Reynolds number,
	ylabel={Maximum Mach number},
	grid=major
      ]
      \addplot table[x index=0,y index=1] {D2Q37A_CM4_4_THM_stability.dat};
      \addplot table[x index=0,y index=2] {D2Q37A_CM4_4_THM_stability.dat};
      \addplot[mark=o] table[x index=0,y index=3] {D2Q37A_CM4_4_THM_stability.dat};
      \addplot[mark=square] table[x index=0,y index=4] {D2Q37A_CM4_4_THM_stability.dat};
      \addplot[mark=x] table[x index=0,y index=5] {D2Q37A_CM4_4_THM_stability.dat};
      \legend{
	{Pr = 0.01},
	{Pr = 0.1},
	{Pr = 1.},
	{Pr = 10.},
	{Pr = 100.}
      }
    \end{semilogxaxis}
  \end{tikzpicture}
  \caption{Stability boundaries in the double shear layer simulation
    using isothermal MRT (top), isothermal CM-MRT (middle) and thermal
    CM-MRT (bottom) models. On the $y$-axis is the maximum Mach number
    ($u_0$) that the simulation can be carried out to $u_0t/Lc =
    2$. For comparison, the Prandtl number is used as a measure of
    $\tau_3$ relative to $\tau_2$ in the top two isothermal cases
    although heat transfer is not simulated there.}
  \label{fg:stabilities}
\end{figure}
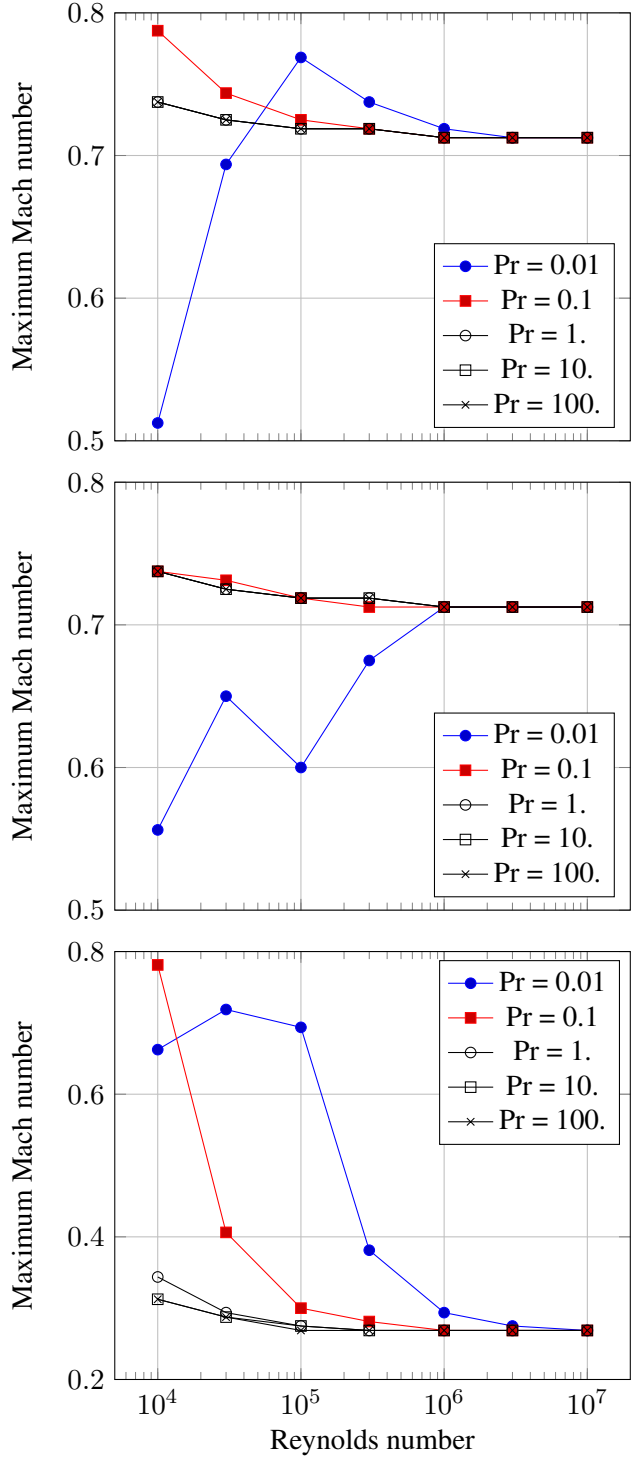

Shown in Fig.~\ref{fg:stabilities} are the stability boundaries in the
$\re$-$\ma$ plane using, from top to bottom, isothermal MRT, isothermal CM-MRT,
and full thermal CM-MRT models. The same $E^9_{2,37}$-A quadrature was used in
all cases and the truncation levels were $M=N=4$. In the isothermal cases, the
temperature field was frozen at unity so that heat transfer is not simulated and
$\tau_3$ becomes a free parameter with no direct impact on the hydrodynamic
equations. To study its effect on numerical stability, the stability boundaries
are plotted for a range of Prandtl numbers defined as $\pr\equiv \nu/\kappa$. It
is evident from Fig.~\ref{fg:stabilities} that for the isothermal simulations
with $\tau_3$ not too far from $\tau_2$, the MRT and CM-MRT perform similarly in
terms of achievable Mach number and Reynolds number. Comparing with the best
result of the regularized LBGK models~\cite{Mattila2017,Coreixas2017}, the
present result ($\ma\sim 0.7$) is approximately 20\% better. For comparison, in
the full thermal case, the maximum achievable Ma number drops to $\sim 0.25$
over a wide range of Prandtl number.  Nevertheless, taking into account that the
grid is severely under-resolved, this maximum is by no means implied as a limit
in practical simulations.

\section{Conclusions and Discussion}

\label{sec:discussion}

In summary, we propose a multiple-relaxation-time collision model by relaxing
the central moments of the distribution function with individually assigned
rates. The collision model is constructed in a way that guarantees that
Chapman-Enskog calculation yields the correct hydrodynamic equation with
separately tunable transport coefficients. Using binomial transform, the central
moments are converted to raw moments for use in lattice Boltzmann models. It is
theoretically shown and numerically verified that viscous and thermal
dissipations are Galilean invariant and mutually independent, allowing a
variable Prandtl number in CFD simulations. The derivation is simple,
lattice-independent and applicable to moments of any order. Excellent numerical
stability was also observed in the double-shear-layer test case.

\begin{acknowledgments}
This work was supported by the National Science Foundation of China Grants
91741101.
\end{acknowledgments}

\appendix

\section{Hermite expansions in the absolute and relative frames}

\label{sec:apdx}

The tensorial Hermite polynomials can be defined by the
\textit{recursive relation}:
\begin{equation}
  \bxi\bH{n}(\bxi) = \bH{n+1}(\bxi) + n\bdelta\bH{n-1}(\bxi),
\end{equation}
where $\bdelta$ is the rank-2 identity tensor.  The first few are:
\begin{subequations}
  \begin{eqnarray}
    \bH{0}(\bxi) &=& 1,\\
    \bH{1}(\bxi) &=& \bxi,\\
    \bH{2}(\bxi) &=& \bxi^2-\bdelta,\\
    \bH{3}(\bxi) &=& \bxi^3- 3\bxi  \bdelta,\\
    \bH{4}(\bxi) &=& \bxi^4- 6\bxi^2\bdelta+3\bdelta^2.
  \end{eqnarray}
\end{subequations}
Inversely, the monomials can be expressed by the Hermite polynomials:
\begin{subequations}
  \label{eq:hermite-r}
  \begin{eqnarray}
    1      &=& \bH{0}(\bxi),\\
    \bxi   &=& \bH{1}(\bxi),\\
    \bxi^2 &=& \bH{2}(\bxi) + \bdelta\bH{0}(\bxi),\\
    \bxi^3 &=& \bH{3}(\bxi) + 3\bdelta\bH{1}(\bxi),\\
    \bxi^4 &=& \bH{4}(\bxi) + 6\bdelta\bH{2}(\bxi) + 3\bdelta^2\bH{0}(\bxi).
  \end{eqnarray}
\end{subequations}

In statistics, the \textit{central} and \textit{raw} moments of a distribution,
defined as the moments about the mean and origin respectively, are related to
each other by the \textit{binomial transform}. Similar relations exist between
the Hermite polynomials and the expansion coefficients in the relative and
absolution reference frames.  First, the following relation can be established
by induction:
\begin{equation}
  \label{eq:trans}
  \bH{n}(\bxi+\bu) = \sum_{i=0}^nC_n^i\bH{i}(\bxi)\bu^{n-i},
\end{equation}
where $C_n^i$ is the \textit{binomial coefficient}. The Hermite polynomials in
the relative and absolute frames are hence related to each other by the
following binomial transforms:
\begin{subequations}
  \label{eq:binomial-transform}
  \begin{eqnarray}
    \bH{n}(\bc) &=& \sum_{i=0}^n(-1)^{n-i}C_n^i\bH{i}(\bxi)\bu^{n-i},\\
    \bH{n}(\bxi) &=& \sum_{i=0}^nC_n^i\bH{i}(\bc)\bu^{n-i},
  \end{eqnarray}
\end{subequations}
where $\bc\equiv\bxi - \bu$.  Let $\ba{n}$ and $\bb{n}$ be respectively the
Hermite coefficients in the absolute and relative frames.  By
Eq.~(\ref{eq:dht-b}), they are related to each other by the binomial transforms:
\begin{subequations}
  \begin{eqnarray}
    \label{eq:trans-a}
    \bb{n} &=& \sum_{i=0}^n(-1)^{n-i}C_n^i\ba{i}\bu^{n-i},\\
    \ba{n} &=& \sum_{i=0}^nC_n^i\bb{i}\bu^{n-i}.
  \end{eqnarray}
\end{subequations}
Explicitly, the leading few expressions are:
\begin{subequations}
  \begin{eqnarray}
    \bb{0} &=& \ba{0},\\
    \bb{1} &=& \ba{1} -  \bu\ba{0},\\
    \bb{2} &=& \ba{2} - 2\bu\ba{1} +  \bu^2\ba{0},\\
    \bb{3} &=& \ba{3} - 3\bu\ba{2} + 3\bu^2\ba{1} -  \bu^3\ba{0},\\
    &\cdots&\nonumber,
  \end{eqnarray}
\end{subequations}
and
\begin{subequations}
  \begin{eqnarray}
    \ba{0} &=& \bb{0},\\
    \ba{1} &=& \bb{1} +  \bu\bb{0},\\
    \ba{2} &=& \bb{2} + 2\bu\bb{1} +  \bu^2\bb{0},\\
    \ba{3} &=& \bb{3} + 3\bu\bb{2} + 3\bu^2\bb{1} + \bu^3\bb{0},\\
    &\cdots&\nonumber.
  \end{eqnarray}
\end{subequations}

%\bibliography{../refs,../misc}
%\bibliographystyle{apsrev4-1}

%merlin.mbs apsrev4-1.bst 2010-07-25 4.21a (PWD, AO, DPC) hacked
%Control: key (0)
%Control: author (72) initials jnrlst
%Control: editor formatted (1) identically to author
%Control: production of article title (-1) disabled
%Control: page (0) single
%Control: year (1) truncated
%Control: production of eprint (0) enabled
%

\end{document}